\documentclass[EPiC]{easychair}

\usepackage{doc}
\usepackage{qtree}
\usepackage{amsfonts}
\usepackage{proving}
\usepackage{booktabs}

\newtheorem{myexample}{Example}

\title{ENIGMA:\\ Efficient Learning-based Inference Guiding Machine}

\author{
   Jan Jakub\r{u}v\inst{1}
\and
    Josef Urban\inst{1}
}

\institute{
  Czech Technical University in Prague,
  Czech Republic,
  \email{\{jakubuv,josef.urban\}@gmail.com}
 }

\authorrunning{Jakub\r{u}v and Urban}

\titlerunning{ENIGMA}

\begin{document}

\maketitle

\begin{abstract}

ENIGMA is a learning-based method for guiding given clause selection in
saturation-based theorem provers.  Clauses from
many proof searches are classified as positive and negative based on their
participation in the proofs.  An efficient classification model
is trained on this data, using fast feature-based characterization of the clauses .  
The learned model is then tightly linked with the core prover and used
as a basis of a new parameterized evaluation heuristic that provides fast ranking of all generated clauses.
The approach is evaluated on the E prover and the CASC 2016 AIM benchmark, showing a large 
increase of E's performance.

\end{abstract}

\setcounter{tocdepth}{2}

\pagestyle{empty}

\section{Introduction: Theorem Proving and Learning}

State-of-the-art resolution/superposition automated theorem provers
(ATPs) such as Vampire \cite{Vampire} and E \cite{Sch02-AICOMM} are today's most advanced tools for
general reasoning across a variety of mathematical
and scientific domains. The stronger the performance of such tools,
the more realistic become tasks such as full computer understanding and
automated development of complicated mathematical theories, and
verification of software, hardware and engineering designs.
While performance of ATPs has steadily grown over the past years due to
a number of human-designed improvements, it is still on average far
behind the performance of trained mathematicians.  Their advanced knowledge-based proof finding
is an enigma, which is unlikely to be deciphered and programmed completely manually in near future.

On large corpora
such as Flyspeck, Mizar and Isabelle, the ATP progress has been mainly
due to learning how to select the most relevant knowledge, based on
many previous proofs \cite{holyhammer,KaliszykU13b,BlanchetteGKKU16,hammers4qed}.  Learning from many proofs
has also recently become a very useful method for automated finding of
parameters of ATP strategies \cite{blistr,JakubuvU17,SchaferS15,KuhlweinU15}, and for learning of sequences of
tactics in interactive theorem provers (ITPs) \cite{GransdenWR15}.
Several experiments with the compact leanCoP \cite{OB03} system
 have recently shown that directly
using trained machine learner for internal clause selection can
significantly prune the search space and solve additional problems \cite{UrbanVS11,KaliszykU15,FarberKU16}. 
An obvious next step is to implement efficient learning-based clause selection also inside the 
strongest superposition-based ATPs. 

In this work, we introduce ENIGMA -- \emph{Efficient learNing-based
Internal Guidance MAchine} for state-of-the-art saturation-based ATPs.  The method applies fast machine learning
algorithms to a large number of proofs, and uses the trained 
classifier together with simpler heuristics to evaluate the millions of clauses generated during the
resolution/superposition proof search.  This way, the theorem prover
automatically takes into account thousands of previous successes and
failures that it has seen in previous problems, similarly to trained humans.
Thanks to a carefully chosen efficient learning/evaluation method and
its tight integration with the core ATP (in our case the E prover),
the penalty for this ubiquitous knowledge-based internal proof
guidance is very low. This in turn very significantly
improves the performance of E in terms of the number of solved problems
in the CASC 2016 AIM benchmark~\cite{Sutcliffe16}.

\section{Preliminaries}

We use $\SetNat$ to denote the set of natural numbers including 0.
When $S$ is a finite set then $|S|$ denotes its size, and $S^n$ where
$n\in\SetNat$ is the $n$-ary Cartesian product of $S$, that is, the set of all
vectors of size $n$ with members from $S$.
When $\mathbf{x}\in S^n$ then we use notation $\mathbf{x}_{[i]}$ to denote its
$i$-th member, counting indexes from 1.
We use $\mathbf{x}^T$ to denote the transposed vector.

Multiset $M$ over a set $S$ is represented by a total function from $S$ to
$\SetNat$, that is, $M(s)$ is the count of $s\in S$ in $M$.
The union $M_1\cup M_2$ of two multisets $M_1$ and $M_2$ over $S$ is the
multiset represented by function $(M_1\cup M_2)(s) = M_1(s)+M_2(s)$ for all
$s\in S$.
We use the notation $\{s_1\mapsto n_1,\ldots,s_k\mapsto n_k\}$ to describe a
multiset, omitting the members with count 0.

We assume a fixed first-order theory with stable symbol names, 
and denote $\Sigma$ its signature, that is, a set of symbols
with assigned arities.
We use $\Lit$ to range over the set of all first-order literals ($\SetLiteral$),
$\Cl$ to range over the set of all first-order clauses ($\SetClause$).
Finally, we use $\Cls$ to range over sets of clauses ($\SetClauses$).

\vspace{-3mm}
\section{Training Clause Classifiers}
\label{sec:training}
There are many different machine learning methods, with different
function spaces they can explore, different training and evaluation
speeds, etc. Based on our previous experiments with premise selection
and with guiding leanCoP, we have decided to choose a very fast and
scalable learning method for the first ENIGMA instantiation.  While
more expressive learning methods usually lead to stronger
single-strategy ATP results, very important aspects of our domain are
that (i) the learning and proving evolve together in a feedback
loop~\cite{US+08-long} where fast learning is useful, and (ii)
combinations of multiple strategies -- which can be provided
by learning in different ways from different proofs -- usually solve
much more problems than the best strategy.
 
After several experiments, we have chosen \LIBLINEAR: open source library 
\cite{DBLP:journals/jmlr/FanCHWL08} for large-scale linear classification. 
This section describes how we use \LIBLINEAR
to train a clause classifier to guide given clause selection.
Section~\ref{sec:examples} describes how training examples can be obtained from
ATP runs. 
Section~\ref{sec:features} describes how clauses are represented as fixed-length
feature vectors.
Finally, Section~\ref{sec:train} describes how to use \LIBLINEAR to train a
clause classifier.

\vspace{-2mm}
\subsection{Extracting Training Examples from ATP Runs}
\label{sec:examples}

Suppose we run a saturation-based ATP to prove a conjecture $\varphi$ in theory $T$.
When the ATP successfully terminates with a proof, we can extract training
examples from this particular proof search as follows.
We collect all the clauses that were selected as given clauses during the
proof search.
From these clauses, those which appear in the final proof are classified as
\emph{positives} while the remaining given clauses as \emph{negative}.
This gives us two sets of clauses, positive clauses $\Clspos$ and negative
clauses $\Clsneg$.

Re-running the proof search 
using the information $(\Clspos,\Clsneg)$ 
to prefer clauses from $\Clspos$ 
as given clauses
should significantly shorten the proof search.
The challenge is to generalize this knowledge to be able to prove new
problems which are in some sense similar.
To achieve that, the positive and negative clauses extracted from proof runs on many related problems are
combined and learned from jointly.

\vspace{-3mm}
\subsection{Encoding Clauses by Features}
\label{sec:features}

In order to use \LIBLINEAR for linear classification (Section~\ref{sec:train}), 
we need to represent clauses as finite \emph{feature vectors}.
For our purposes, a feature vector $\mathbf{x}$ representing a clause $C$ is a
fixed-length vector of natural numbers whose $i$-th member $\mathbf{x}_{[i]}$
specifies how often the $i$-th feature appears in the clause $C$.

Several choices of clause features are possible~\cite{ckjujv-ijcai15}, for example sub-terms, their generalizations,
or paths in term trees.
In this work we use term walks of length 3 as follows.
First we construct a feature vector for every literal $\Lit$ in the clause $\Cl$.
We write the literal $\Lit$ as a tree where nodes are labeled by the symbols from
$\Sigma$.
In order to deal with possibly infinite number of variables and Skolem
symbols, we substitute all variables and Skolem symbols with special
symbols.
We count for each triple of symbols $(s_1,s_2,s_3)\in\Sigma^3$, the
number of directed node paths of length 3 in the literal tree, provided the trees
are oriented from the root.
Finally, to construct the feature vector of clause $\Cl$, we sum the vectors of all
literals $\Lit\in\Cl$.

More formally as follows.
We consider a fixed theory $T$, hence we have a fixed signature $\Sigma$.
We extend $\Sigma$ with 4 special symbols for variables ($\oasterisk$), Skolem
symbols ($\odot$), positive literals ($\oplus$), and negative literals ($\ominus$).
A \emph{feature} $\Feature$ is a triple of symbols from $\Sigma$. The set of
all features is denoted $\SetFeature$, that is, $\SetFeature=\Sigma^3$.
\emph{Clause (or literal) features} $\Features$ is a multiset of features, thus recording
for each feature how many times it appears in a literal or a clause.
We use $\Features$ to range over literal/clause features and the set of all
literal/clause features (that is, feature multisets) is denoted $\SetFeatures$.
Recall that we represent multisets as total functions from $\SetFeature$ to
$\SetNat$.
Hence every member $\Features\in\SetFeatures$ is a total function of the type
``$\SetFeatures\rightarrow\SetNat$'' and we can write $\Features(\Feature)$ to
denote the count of $\Feature$ in $\Features$.

Now it is easy to define function $\SYMfeatures$ of the type 
   ``$\SetLiteral\rightarrow\SetFeatures$''
which extracts features $\Features$ from a literal $\Lit$.
For a literal $\Lit$, we construct a rooted \emph{feature tree} with nodes
labeled by the symbols from $\Sigma$.
The feature tree basically corresponds to the tree representing literal $\Lit$ with the
following exceptions.
The root node of the tree is labeled by $\sympos$ iff $\Lit$ is a positive
literal, otherwise it is labeled by $\symneg$.
Furthermore, all variable nodes are labeled by the symbol $\symvar$ and all nodes
corresponding to Skolem symbols are labeled by the symbol $\symsko$.

\begin{figure}
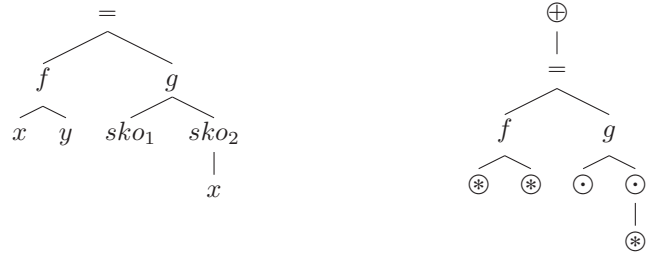

\Tree [.$=$ [.$f$ $x$ $y$ ] [.$g$ $sko_1$  [.$sko_2$ $x$ ] ] ]
\Tree [.$\sympos$ [.$=$ [.$f$ $\symvar$ $\symvar$ ] [.$g$ $\symsko$  [.$\symsko$ $\symvar$ ] ] ] ]
\caption{The tree representing literal $\Lit_1$ from
Example~\ref{ex:feature-tree} (left) and its corresponding feature tree
(right).}
\label{fig:features}
\end{figure}

\begin{myexample}
\label{ex:feature-tree}
Consider the following equality literal $\Lit_1:
   f(x,y)=g(sko_1,sko_2(x))
$ 
with Skolem symbols $\text{sko}_1$ and
$\text{sko}_2$, and with variables $x$ and $y$.
In Figure~\ref{fig:features}, the tree representation of $\Lit_1$ is depicted on the left, while the
corresponding feature tree used to collect features is shown on the right.
\hfill\qed

\end{myexample}

Function $\SYMfeatures$ constructs the feature tree of a literal $\Lit$ and
collects all directed paths of length 3.
It returns the result as a feature multiset $\Features$.

\begin{myexample}

For literal $L_2 : P(x)$ we obtain $\features{L_2}=\{\fea{\sympos}{P}{\symvar}\mapsto 1\}$.
For literal $L_3 : \lnot Q(x,y) $ we have  $\features{\lnot
Q(x,y)}=\{(\symneg,Q,\symvar)\mapsto 2\}$.
Finally, for literal $L_1$ from Example~\ref{ex:feature-tree} we obtain the
following multiset.
\[\{\quad
   \fea{\sympos}{=}{f}\mapsto 1\ ,\quad
   \fea{\sympos}{=}{g}\mapsto 1\ ,\quad
   \fea{=}{f}{\symvar}\mapsto 2\ ,\quad
   \fea{=}{g}{\symsko}\mapsto 2\ ,\quad
   \fea{g}{\symsko}{\symvar}\mapsto 1
\quad\}\]
\hfill\qed
\end{myexample}

Finally, the function $\SYMfeatures$ is extended to clauses 
($
   \SYMfeatures : \SetClause \rightarrow \SetFeatures
$)
by multiset union as 
$
   \features{\Cl} = \bigcup_{\Lit\in\Cl}\features{\Lit}
$.

\subsubsection{A Technical Note on Feature Vector Representation}

In order to use \LIBLINEAR, we transform the feature multiset
$\Features$ to a vector of numbers of length $|\SetFeature|$.
We assign a natural index to every feature and we construct a vector whose
$i$-th member contains the count $\Features(\Feature)$ where $i$ is the index of
feature $\Feature$.
Technically, we construct a bijection $\SYMsym$ between $\Sig$ and
$\{0,\ldots,|\Sig|-1\}$ which encodes symbols by natural numbers.
Then we construct a bijection between $\SetFeature$ and
$\{1,\ldots,|\SetFeature|\}$ which encodes features by numbers\footnote{
We use
$
   \code{\phi} =
   \sym{\phi_{[1]}}\cdot|\Sigma|^2+\sym{\phi_{[2]}}\cdot|\Sigma|+\sym{\phi_{[3]}}+1
$.
}.
Now it is easy to construct a function $\SYMvector$ which translates $\Phi$ to
a vector from $\SetNat^{|\SetFeature|}$ as follows:
\[
   \vector{\Phi} = \mathbf{x} \mbox{ such that } \mathbf{x}_{[\code{\phi}]} = \Phi(\phi) \mbox{ for
   all }\Feature\in\SetFeature
\]

\subsection{Training Clause Classifiers with \LIBLINEAR}
\label{sec:train}

Once we have the training examples $(\Clspos,\Clsneg)$ and encoding of clauses by
feature vectors, we can use \LIBLINEAR to construct a classification model.
\LIBLINEAR implements the function $\SYMtrain$
of the type
   ``$\SetClauses \times \SetClauses \rightarrow \SetModel$''
which takes two sets of clauses
(positive and negative examples) and constructs a classification model.
Once we have a classification model
   $\Model = \train{\Cls_\clspos,\Cls_\clsneg}$, \LIBLINEAR provides a function
$\SYMpredict$ 
of the type
   ``$\SetClause \times \SetModel\rightarrow\{\clspos,\clsneg\}$''
which can be used to predict clause classification as positive
($\clspos$) or negative ($\clsneg$).

\LIBLINEAR supports several classification methods, but we have so far
used only the default solver L2-SVM (L2-regularized L2-loss Support Vector
Classification)~\cite{DBLP:conf/colt/BoserGV92}.
Using the functions from the previous section, we can translate the training examples $(\Clspos,\Clsneg)$ 
to the set of instance-label pairs $(\mathbf{x}_i,y_i)$, where 
$i\in\{1,\ldots,|\Clspos|+|\Clsneg|\}$, 
$\mathbf{x}_i\in\SetNat^{|\SetFeature|}$, 
$y_i\in\{\clspos,\clsneg\}$.
A training clause $\Cl_i$ is translated to the feature vector
$\mathbf{x}_i=\vector{\features{\Cl_i}}$ and the corresponding $y_i$
is set to $\clspos$ if $\Cl_i\in\Clspos$ or to $\clsneg$ if $\Cl_i\in\Clsneg$.
Then, \LIBLINEAR solves the following optimization problem:
\[
   \min_{\textbf{w}} \big( \frac{1}{2}\textbf{w}^T\textbf{w} +
   c\sum_{i=1}^l\xi(\mathbf{w},\mathbf{x_i},y_i) \big)
\]
for $\mathbf{w}\in\SetReal^{|\SetFeature|}$, where
$c>0$ is a penalty parameter and $\xi$ is the following loss function.
\[
   \xi(\mathbf{w},\mathbf{x_i},y_i) = \max(1-y'_i\mathbf{w}^T\mathbf{x_i},0)^2
   \mbox{ where } y'_i = 
   \begin{cases}
      1  & \mbox{iff } y_1 = \clspos \\
      -1 & \mbox{otherwise}
   \end{cases}
\]
\LIBLINEAR implements a coordinate descend
method~\cite{DBLP:conf/icml/HsiehCLKS08} and a trust region Newton
method~\cite{DBLP:journals/jmlr/LinWK08}.

The model computed by \LIBLINEAR is basically the vector $\textbf{w}$ obtained by
solving the above optimization problem.
When computing the prediction for a clause $\Cl$, the clause is translated to the
corresponding feature vector $\textbf{x}=\vector{\features{\Cl}}$ and \LIBLINEAR classifies $\Cl$ as
positive ($\clspos$) iff $\textbf{w}^T\textbf{x}>0$.
Hence we see that the prediction can be computed in time
$O(\max(|\SetFeature|,\len{\Cl}))$ where $\len{\Cl}$ is the length of clause
$\Cl$ (number of symbols).

\vspace{-2mm}
\section{Guiding the Proof Search}
\label{sec:guide}

Once we have a \LIBLINEAR model (classifier) $\Model$, we construct a clause weight
function which can be used inside the ATP given-clause loop to evaluate the generated clauses.
As usual, clauses with smaller weight are selected before those with a higher
weight.
First, we define the function $\SYMpredict$ which assigns a smaller number to positively
classified clauses as follows:
\[
   \preweight{\Cl,\Model} = \begin{cases}
      1  & \mbox{iff } \predict{\Cl,\Model}=\clspos
   \\ 10 & \mbox{otherwise}
   \end{cases}
\]
In order to additionally prefer smaller clauses to larger ones, we add the
clause length to the above predicted weight.
We use $\len{\Cl}$ to denote the length of $C$ counted as the number of symbols.
Furthermore, we use a real-valued parameter $\gamma$ to multiply the length as
follows.
\[
   \weight{\Cl,\Model} = \gamma\cdot\len{\Cl} + \preweight{\Cl,\Model}
\]

This scheme is designed for the E automated prover which uses
\emph{clause evaluation functions} (CEFs) to select the given clause.
A clause evaluation function $\textit{CEF}$ is a function which assigns a real
weight to a clause.
The unprocessed clause with the smallest weight is chosen to be the given clause.
E allows combining several CEFs to jointly guide the 
proof search.
This is done by specifying a finite number of CEFs together with their \emph{frequencies} as follows:
$
   (f_1\star\mathit{CEF}_1,\ldots,f_k\star\mathit{CEF}_k)
$.
Each frequency $f_i$ denotes how often the corresponding $\mathit{CEF}_i$
is used to select a given clause in this weighted round-robin scheme.
We have implemented learning-based guidance as a new CEF
given by the above $\SYMweight$ function.
We can either use this new CEF alone or combine it with other CEFs already defined in
E.

\vspace{-2mm}
\section{Experimental Evaluation}

We use the AIM\footnote{AIM is a long-term project on proving open algebraic conjectures by Kinyon and Veroff.} category of the CASC 2016 competition for evaluation.
This benchmark fits our needs as it targets internal guidance in ATPs
based on training and testing examples.
Before the competition, 1020 training problems were provided for the training of
ATPs, while additional 200 problems were used in the competition.
Prover9 proofs were provided along with all the training problems.
Due to several interesting issues,\footnote{E.g., different term orderings, rewriting settings, 
etc., may largely change the proof search.}
we have decided not to use the training Prover9 proofs yet and instead 
find as many proofs as possible by a single E strategy.

Using fast preliminary evaluation, we have selected a strong E\footnote{We use E 1.9 and Intel Xeon 2.6GHz workstation for all experiments.} strategy $S_0$ (see Appendix~\ref{sec:strat}) which
can by itself solve 239 training problems with a 30 s timeout.
For comparison, E's auto-schedule mode (using optimized strategy scheduling) 
can solve 261 problems.
We train a clause classifier model $\Model_0$ (Section~\ref{sec:training}) on the 239 proofs 
and then run
E enhanced with the
classifier $\Model_0$ in different ways to
obtain even more training examples.
Either we use the classifier CEF based on $\Model_0$ (i.e., function $\SYMweight(C,\Model_0)$ from
Section~\ref{sec:guide}) alone, or combine it with the CEFs
from $S_0$ by adding $\SYMweight(C,\Model_0)$ to $S_0$
with a grid of frequencies ranging over
\{1,5,6,7,8,9,10,15,20,30,40,50\}.  Furthermore, every combination may
be run with a different value of the parameter
$\gamma\in\{0,0.1,0.2,0.4,0.7,1,2,4,8\}$ of the function
$\SYMweight(C,\Model_0)$.  All the methods are run with 30 seconds time
limit, leading to the total of 337 solved training problems. As
expected, the numbers of processed clauses and the solving times on
the previously solved problems are typically very significantly
decreased when using $\SYMweight(C,\Model_0)$. This is a good sign, however, the ultimate test of ENIGMA's
capability to learn and generalize is to evaluate the trained
strategies on the testing problems. This is done as follows.

On the 337 solved training problems, we (greedily) find 
that 4 strategies are needed to cover the whole set.
The strongest strategy is our classifier $\SYMweight(C,\Model_0)$ alone with $\gamma=0.2$, solving 318 problems.
Another 15 problems are added by combining $S_0$ with the trained classifier 
using frequency 50 and $\gamma=0.2$.
Three problems are contributed by $S_0$ and two by the trained classifier alone
using $\gamma=0$. We take these four strategies and use only the proofs they found to train a new
enhanced classifier $\Model_1$.
The proofs yield 6821 positive and 219012 negative examples.
Training of $\Model_1$ by \LIBLINEAR takes about 7 seconds -- 2 seconds
for feature extraction and 5 seconds for learning.
The classifier evaluation on the training examples takes about 6 seconds and
reaches 97.6\% accuracy (ratio of the correctly classified clauses).

This means that both the feature generation and the model evaluation
times per clause are at the order of 10 microseconds. This is
comparable to the speed at which clauses are generated by E on our
hardware and evaluated by its built-in heuristics. Our learning-based
guidance can thus be quickly trained and used by normal users of E,
without expensive training phase or using multiple CPUs or GPUs
for clause evaluation.

Then we use the $\Model_1$ classifier to attack the 200 competition problems using 180 s time limit
as in CASC.
We again run several strategies: both $\SYMweight(C,\Model_1)$ alone and
combined with  $S_0$ with different frequencies and parameters $\gamma$.
All the strategies solve together 52 problems and only 3 of the strategies are needed for this.
While $S_0$ solves only 22 of the competition problems, our strongest strategy solves 41 problems, see Table~\ref{Table1}. 
This strategy combines $S_0$ with $\SYMweight(C,\Model_1)$  using frequency
30 and $\gamma=0.2$.
7 more problems are contributed by $\SYMweight(C,\Model_1)$  alone with $\gamma=0.2$ and 4 more problems 
are added by the E auto-schedule mode.
For comparison, Vampire solves 47 problems (compared to our 52 proofs) in 3*180
seconds per problem (simulating 3 runs of the best strategies, each for 180 seconds).

\begin{table}[h!]
\begin{small}

\centering
\begin{tabular}{cccccccccc}
\toprule
      & auto-schedule & 0 & 1 & 5 & 10 & 15 & 30 & 50 & $\infty$ \\\midrule
0    &29 & 22& - &-   & -   & -   & \bf{18} & 17 &16\\
0.2 & - & -&23&31&32&40 & \bf{41} & 33 &40\\
8    & - & -&23&31&31&40 & \bf{41} & 33& 35\\\bottomrule
\end{tabular}
\caption{\label{Table1}\small{Performance of the differently parameterized (frequency and $\gamma$ of $\SYMweight(C,\Model_1)$ combined with $S_0$)  trained evaluation heuristics on the 200 AIM CASC 2016 competition problems. Frequency $0$ (third column) for $\SYMweight(C,\Model_1)$ means that $S_0$ is used alone, whereas $\infty$ means that $\SYMweight(C,\Model_1)$ is used alone. The empty entries were not run.}} 
\end{small}
\end{table}

\vspace{-6mm}
\subsection{Looping and Boosting}
The recent work on the premise-selection task has shown that
typically there is not a single optimal way how to guide  proof
search. Re-learning from new proofs as introduced by MaLARea and
combining proofs and learners
usually outperforms a
single method.  Since we are using a very fast classifier here, we can
easily experiment with giving it more and different data.

First such experiment is done as follows. 
We add the proofs obtained on the solved 52 competition problems to
the training data obtained from the 337 solved training
problems. Instead of immediately re-learning and re-running (as in the
MaLARea loop), we however first boost all positive examples (i.e.,
clauses participating in the proofs) by repeating them ten times in the
training data. This way, we inform the learner to more strongly avoid
misclassifying the positive examples as negative, than the other way
round. The resulting clasifier $\Model_2$ has lower overall accuracy on all of
the data (93\% vs. 98\% for the unboosted), however, its accuracy on
the relatively rare positive data grows significantly, from 12.5\% to 81.8\%. 

Running the most successful strategy using $\Model_2$ instead of $\Model_1$ indeed
helps.  In 180 s, it solves additional 5 problems (4 of them not
solved by Vampire), all of them in less than 45 s.  This raises
ENIGMA's performance on the competition problems to 57 problems (in
general in 600 s).
Interestingly, the
second most useful strategy (now using $\Model_2$ instead of $\Model_1$) which is much more focused on doing
inferences on the positively classified clauses, solves only two of
these new problems, but six times faster.  It is clear that we can
continue experimenting this way with ENIGMA for long time, producing
quickly a large number of strategies that have quite different search
properties. In total we have proved 16 problems unsolved by Vampire.

\section{Conclusions}

The first experiments with ENIGMA are extremely encouraging. While the
recent work on premise selection and on internal guidance
for leanCoP indicated that large improvements are possible, this is
the first practical and usable improvement of a state-of-the-art ATP
by internal learning-based guidance on a large CASC benchmark.  It is
clear that a wide range of future improvements are possible: the learning
could be dynamically used also during the proof search, training problems selected according to their similarity with the current problem,\footnote{In an initial experiment, a simple nearest-neighbor selection of training problems for the learning further decreases the solving times and proves one more AIM problem unsolved by Prover9.} more sophisticated learning and feature characterization methods could be employed, etc.

The
magnitude of the improvement is unusually big for the ATP field, and
similar to the improvements obtained with high-level learning in
MaLARea 0.5 over E-LTB (sharing the same underlying engine) in CASC
2013~\cite{malar14}. We believe that this may well mark the arrival
of ENIGMAs -- efficient learning-based
inference guiding machines -- to the automated reasoning, as crucial and indispensable technology for building the strongest automated theorem provers.

\vspace{-2mm}
\section{Acknowledgments}

We thank Stephan Schulz for his open and modular implementation of E and its many features
that allowed us to do this work. We also thank the Machine Learning Group at National Taiwan University
for making \LIBLINEAR openly available.
This work was supported by the ERC Consolidator grant no.\ 649043 \textit{AI4REASON}.

\vspace{-3mm}
\bibliographystyle{abbrv}
\bibliography{learning,ate11}

\appendix

\section{The E Prover Strategy Used in Experiments}
\label{sec:strat}

The following fixed E strategy $S_0$, described by its command line arguments, was
used in the experiments:

\begin{verbatim}
--definitional-cnf=24 --destructive-er-aggressive --destructive-er 
--prefer-initial-clauses -F1 --delete-bad-limit=150000000 --forward-context-sr 
-winvfreqrank -c1 -Ginvfreq -WSelectComplexG --oriented-simul-paramod -tKBO6 
-H(1*ConjectureRelativeSymbolWeight(SimulateSOS,0.5,100,100,100,100,1.5,1.5,1),
   4*ConjectureRelativeSymbolWeight(ConstPrio,0.1,100,100,100,100,1.5,1.5,1.5),
   1*FIFOWeight(PreferProcessed),
   1*ConjectureRelativeSymbolWeight(PreferNonGoals,0.5,100,100,100,100,1.5,1.5,1),
   4*Refinedweight(SimulateSOS,3,2,2,1.5,2))
\end{verbatim}

\end{document}